\documentclass[conference]{IEEEtran}
\IEEEoverridecommandlockouts
\usepackage{cite}
\usepackage{amsmath,amssymb,amsfonts}
\usepackage{algorithmic}
\usepackage{graphicx}
\usepackage{textcomp}
\usepackage{xcolor}
\usepackage{booktabs}   
\usepackage{multirow}
\usepackage{balance}
\usepackage[colorlinks,
            linkcolor=black,
            anchorcolor=black,
            citecolor=black]{hyperref}

\def\BibTeX{{\rm B\kern-.05em{\sc i\kern-.025em b}\kern-.08em
    T\kern-.1667em\lower.7ex\hbox{E}\kern-.125emX}}
\begin{document}

\title{Passive Underwater Acoustic Signal Separation based on Feature Decoupling Dual-path Network\\

}

\author{Yucheng Liu and Longyu Jiang

\thanks{Manuscript created March, 2025; 
This work was supported by the National Natural Science Foundation of China under Grant 61871124, and developed by the Smart Ocean Information Processing Laboratory, Southeast University.
(\textit{Corresponding author: Yucheng Liu.})  

Yucheng Liu and Longyu Jiang are with the School of Computer Science and Engineering, Southeast University (SEU), Nanjing 210000, China (e-mail: 220224322@seu.edu.cn; JLY@seu.edu.cn).}}

\maketitle

\begin{abstract}
Signal separation in the passive underwater acoustic domain has heavily relied on deep learning techniques to isolate ship radiated noise. 
However, the separation networks commonly used in this domain stem from speech separation applications and may not fully consider the unique aspects of underwater acoustics beforehand,
such as the influence of different propagation media, signal frequencies and modulation characteristics.
This oversight highlights the need for tailored approaches that account for the specific characteristics of underwater sound propagation.
This study introduces a novel temporal network designed to separate ship radiated noise by employing a dual-path model and a feature decoupling approach. 
The mixed signals' features are transformed into a space where they exhibit greater independence, with each dimension's significance decoupled. Subsequently, a fusion of local and global attention mechanisms is employed in the separation layer. 
Extensive comparisons showcase the effectiveness of this method when compared to other prevalent network models, as evidenced by its performance in the ShipsEar and DeepShip datasets.
\end{abstract}


\begin{figure*}[htbp]
    \centerline{\includegraphics[width=1.0\linewidth]{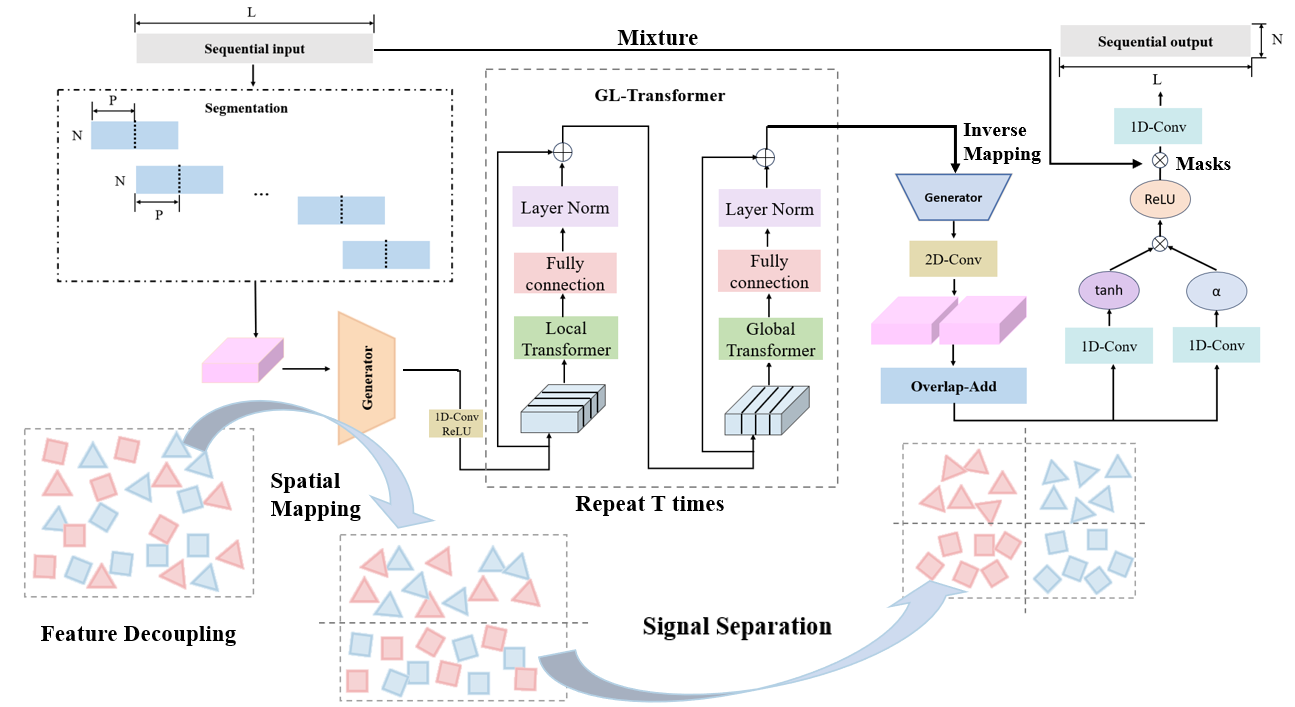}}
    \caption{Network Architecture Overview. Two types of mixed signals are distinguished by red and blue, and different features are represented by triangles and rectangles. Feature decoupling is performed first, and then signal separation is performed through the architecture based on the proposed Multi-scale Dual-path Transformer.}
    \label{fig}
\end{figure*}

\section{Introduction}
\IEEEPARstart{U}{nderwater} acoustic signals play a crucial role in marine operations, especially in the utilization of passive sonar systems for receiving and analyzing signals.
Passive underwater targets encompass a range of sources, including marine organisms using sound for communication and navigation, ship radiated noise from civilian and military vessels, and natural environmental sounds like waves and wind. 
Separating these passive underwater targets is especially focused on effectively distinguishing mixed ship radiated noise. However, the complex underwater environment presents challenges due to the presence of multiple noise sources and reverberations, 
making the separation of ship radiated noise a significant undertaking.

Conventional techniques for underwater acoustic separation are categorized as single-channel-based methods and multi-channel array-based separation techniques.
Single-channel filtering methods include spectral subtraction, Wiener filtering, and more. 
Multi-channel methods include subspace-based methods and blind source separation. Spectral subtraction \cite{b1} is a technique initially used for speech enhancement, 
estimating the background noise spectrum by computing the average magnitude or energy spectrum of the mixed signal in the early frames. 
For non-stationary noise like ship radiated noise, whose characteristics may vary over time, spectral subtraction may not be suitable. 
Wiener filtering \cite{b2} is a commonly used signal separation method that minimizes the square difference between the output and the desired signal by solving a Toeplitz matrix. 
Adaptive filtering \cite{b3, b4} is based on linear filtering techniques like Wiener filtering and Kalman filtering, allowing real-time parameter adjustments, 
suitable for processing dynamic and non-stationary signals.

Additionally, there are subspace-based methods \cite{b5}  involving the construction of a model of the signal subspace and extracting underlying sources from the observed mixtures using techniques like singular value decomposition (SVD) \cite{b6, b7, b8}  or principal component analysis (PCA) \cite{b9} . 
It relies on assumptions about the characteristics of the signal and noise, deviating from these assumptions may lead to a decrease in separation performance.

In the field of signal separation, blind source separation based on Independent Component Analysis (ICA) has been widely applied \cite{b10, b11}. BSS exploits the statistical independence or different statistical properties of source signals to separate mixed signals.  
Gaeta et al. \cite{b12} firstly used BSS to estimate the impulse response function of underwater channels. 
Kirstein \cite{b13} investigated the effects of sea surface multipath on synthetic aperture sonar using BSS. 
Kamal et al. \cite{b14} combined slow feature analysis with BSS for underwater acoustic signals. In 2015, Tu et al. \cite{b15} separated underwater acoustic signals based on the negentropy FastICA algorithm. 
Li et al. \cite{b16} used spatial filters with a hydrophone array to separate underwater sources.
However, BSS requires the number of observed audio signals to be greater than or equal to the number of sources due to the statistical
independence between sources. 

In recent years, deep learning-based signal separation techniques typically employ end-to-end approaches, taking mixed signals in the time domain or time-frequency domain as input. 
Research on convolutional neural networks \cite{b17}, u-net \cite{b18}, and other deep learning networks has received widespread attention. 
The Deep Complex UNet (DCUNet) \cite{b19} combines the advantages of deep complex networks and UNet by estimating Complex Ratio Masks (CRMs) to handle complex spectrograms. 
The Residual u-net (Res-UNet) \cite{b20} is commonly used for sound extraction in music, utilizing Complex Ideal Ratio Masks (CIRMs) to address challenges in CIRM estimation due to the sensitivity of the real and imaginary parts of the complex mask to signal time shifts. 

In recent years, the focus of signal separation in underwater acoustic research has shifted towards a data-driven approach, 
leveraging features as the foundation and utilizing deep learning models for learning. 
Unlike speech separation, underwater acoustic signals face challenges due to their unique propagation medium 
and frequency range, leading to issues such as time variance and multipath effects. 
Simply expanding networks in terms of depth and breadth can encounter limitations in this context.
This paper proposes a time-domain separation network model based on the decoupling of mixed signal features and 
the GL-transformer. The main contributions are as follows:

1) For the reshaped three-dimensional tensor, we perform feature decoupling, mapping it to another feature space to relatively separate the features of mixed signals, hence named Indiformer. 

2) In our model, an improved transformer named GL-Transformer is used to group the spatial dimensions of the features, calculate attention within each group, and finally fuse local attention and global attention. Separation validation experiments under various conditions demonstrate that the network's separation results are more accurate and reliable.

\section{Related work}

Recurrent Neural Networks (RNN) \cite{b21} can learn correlations between signal features by processing long sequences with recursive connections. 
Long Short-Term Memory (LSTM) networks \cite{b22} optimize for vanishing and exploding gradients during training.

Most deep learning-based audio signal separation techniques utilize complex masking models based on time-frequency spectrograms. 
This involves transforming the time-domain waveform into time-frequency spectrograms. 
Directly performing audio signal separation in the time domain is an important approach. 
Time-domain audio separation networks (TasNet) \cite{b23} and convolutional time-domain audio separation networks (Conv-TasNet) \cite{b24} estimate masks directly from the time-domain waveform, preserving phase information and reducing network size through one-dimensional convolutions. 
In TasNet, signals are directly separated in the time domain using an encoder-decoder framework. By skipping the frequency decomposition step, it simplifies the separation task to estimating speech masks on the encoder outputs, which are then synthesized by the decoder. This approach of estimating weights corresponding to each source from the mixed signal has since found widespread use in time-domain speech separation.
Conv-TasNet follows a separation approach where a weight mask is applied to the encoder's output, and the modified representation is used to generate audio through a linear decoder. Within its time convolution network, an initial 512-dimensional vector is processed pairwise to compute a new vector using a convolutional kernel. Subsequent convolutional layers operate with increasing gaps, with each layer doubling the gap size, known as the dilation factor, leading to exponential growth. With more layers, the resulting mask can cover features of more sample points effectively.

Dual-path recurrent neural network (DPRNN) \cite{b25} is also a time-domain-based signal separation method. 
The model reshapes audio sequences into blocks and employs two paths, intra-chunk RNN and inter-chunk RNN, to learn intra-block and inter-block relationships and separate by estimating masks. 
It improves the performance of single-channel audio separation algorithms using dual-path architecture when dealing with long mixed audio sequences. 
It achieves this by breaking down the input mixed audio sequence into blocks and iteratively modeling within-block and across-block information, thereby learning both local and global features, which enhances the separation performance effectively.
Mossformer \cite{b26} effectively addresses indirect element interactions across blocks in the dual-path architecture, proposing a transformer-based speech separation model architecture with joint self-attention and gated single-head mechanisms.

\begin{figure*}[htbp]
    \centerline{\includegraphics[width=0.93\linewidth]{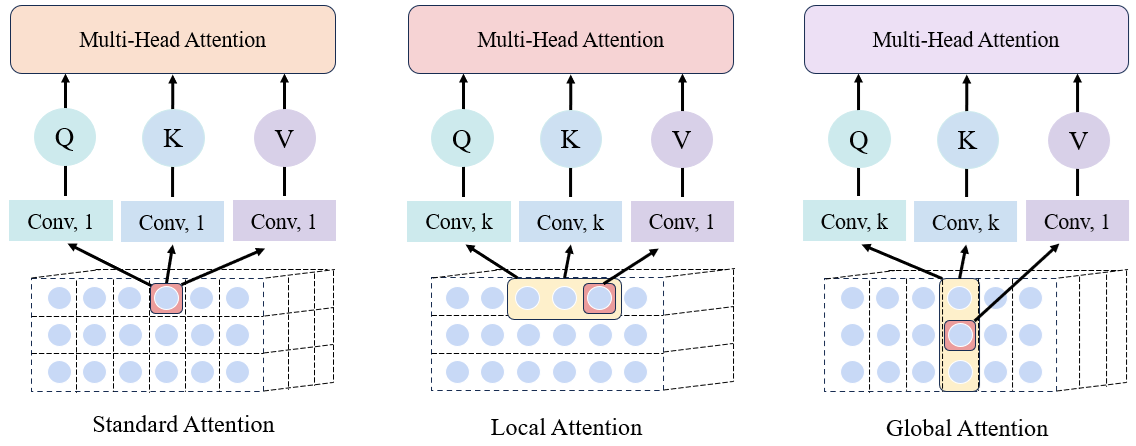}}
    \caption{The computation strategy for \textit{Local and Global Attention} in GL-Transformer.}
    \label{fig}
\end{figure*}

\section{Method}

\subsection{Dual-Path Architecture}
The core idea behind the dual-path architecture involves transforming a lengthy sequence into a cubic tensor made up of multiple blocks and sequentially processing this tensor within and between blocks \cite{b25}. This process consists of three stages: Segmentation, Block Processing, and Overlap-Add.

Segmentation aims to divide a long sequence into smaller segments, which are then horizontally combined to create a stacked block. It's important to note that these adjacent blocks have overlapping parts. Each segment has a length of $k$, with an overlapping part of length $p$. When $k=2p$, each adjacent block precisely overlaps half with the preceding block and half with the succeeding block. 
Assuming the original sequence features are n-dimensional and divided into $s$ blocks, a tensor of size $n \times k \times s$ is obtained. 
The advantage of forming a tensor is that it inherently contains many sampling points, eliminating the need to gradually expand the receptive field layer by layer through convolutions.

In the Block Processing, two types of processing are sequentially applied to the tensor obtained from the previous step. Intra-chunk Transformer processes each segment from the Segmentation step individually, focusing on the internal features of each segment. On the other hand, the inter-chunk Transformer extracts the sampling points at the same coordinates in each segment for processing, aiming to capture the relationships between different segments. Intra-chunk processing learns features of local neighboring regions, while inter-chunk processing learns the connections between different segments. 
The dual-path structure consolidates the segmented sequence for subsequent feature learning at different scales.

Once all the steps are completed, the compressed tensor in block form needs to be unfolded and reshaped back into a long sequence. Since the size of the tensor $n \times k \times s$, remains unchanged throughout the entire model, the reshaped form will still be a sequence of length $L$ with $N$ dimensions.

\subsection{Our Improved Dual-path Separation Network}

The network operates on the dual-path architecture as well. 
Initially, the input sequence is transformed into a block and spatially arranged in the generator. 
Following this, the features of distinct dimensions within the sequence are disentangled. 
This is illustrated in Figure 1, where triangles and rectangles denote the separation of different features, while signals of different colors remain mixed. 
As the blocks traverse the GL-Transformer, they are assembled and attention is computed within and across these groups, ultimately leading to the segregation of distinct signals. 
Subsequently, the feature tensor is remapped to its original space, and the mask for each source signal is learned through two-dimensional convolution. 
Overlap-Add is employed to restore the original shape, and ultimately, the separated time series is reconstructed by element-wise multiplication between the mixed audio and the mask.

\textbf{Reversible feature decoupling module. }
This step involves transforming the obtained data into a space where the features are more independent. 
The goal is to ensure that the features in this space are not dependent on each other. 
Before separating the signals, we first isolate all the features within the mixed signals. 
To achieve this, we employ a generator ${G^*}$ to map the data into a new representation space using maximum likelihood estimation, 
where ${x_1, x_2,..., x_m}$ represent samples from the actual data distribution:

\begin{equation}
    G^*=\mathop{\arg\max}\limits_{G}\sum_{i = 1}^{m}logP_G(x^i)\label{eq}
\end{equation}

The likelihood function can be calculated as follows:

\begin{equation}
    P_G(x^i)=\pi(z^i)|det(J_{G^{-1}})|\label{eq}
\end{equation}

The generated distribution $x$ can be mapped to the initial distribution $z$ through the inverse process of $G^*$, 
which can be used to train $G^{-1}$. After taking the logarithm, 
the following expression can be obtained:

\begin{equation}
    logP_G(x^i)=log\pi(G^{-1}(x^i))+log|det(J_{G^{-1}})|\label{eq}
\end{equation}

When dealing with the variable $G^{-1}$, it's essential to train $G^{-1}$ to maximize the likelihood function. 
However, calculating the Jacobian determinant $det J_{G ^ {-1}}$ 
is challenging in practice. 
Let $x{\in}X$, where x is divided into two segments $x1$ and $x2$, 
yielding the definition of $y=(y1, y2)$, where:

\begin{equation}  
    \left\{  
         \begin{array}{lr}  
         y_1=x_1 &   \\ 
         y_2=G^*(x_2;l(x_1)) \\ 
         \end{array}  
    \right.  
    \end{equation} 

Among them, $l$ represents a fully connected layer. 
The Jacobian determinant of $y$ with respect to $x$ is as follows:

\begin{equation}
    \frac{ \partial y }{ \partial x } = 
        \begin{bmatrix}
            \vspace{1.5ex}
            I & 0 \\ 
            \vspace{1.5ex}
            \dfrac{\partial y_2}{\partial x_1} & \dfrac{\partial y_2}{\partial x_2}
        \end{bmatrix}
    \end{equation}

In this scenario, the matrix has a clear structure: 
the top-left part is the identity matrix, the top-right part is all zeros, and the bottom-right part is a diagonal matrix. This arrangement simplifies the calculation process. Furthermore, because $G^*$ is reversible, 
we can employ this approach to reestablish tensors to their initial feature space.

Typically, we collect data points $x^i$ from the actual distribution to train the reverse process $G^-1$ of $G^*$. 
Afterward, we select a point $z^j$ from $z$ and produce a sample $x^j=G(z^j)$, 
representing the distribution of independent features derived from the mapping process.

\begin{table*}[htbp]
    \renewcommand{\arraystretch}{1.2}
    \setlength{\tabcolsep}{25pt}
    \caption{The model parameters configuration}
    \begin{center}
        \scalebox{1.25}{
        \begin{tabular}{ccc}
            \toprule
            \textbf{Parameter} & \textbf{Parameter Description} & \textbf{Value} \\
            \midrule
            n\_src & Number of masks to estimate & 2 \\
            chunk\_size & Window size of overlap and add processing & 100 \\
            hop\_size & Hop size of overlap and add processing & 50 \\
            n\_repeat & Number of repetitions of the dual path structure & 6 \\
            n\_head & Number of heads for multi-head attention & 4 \\
            dropout & proportion of discarded neurons & 0.1 \\
            $k$ & Number of convolution kernels in separation layer & 128 \\
            $l$ & Convolutional kernel size in separation layer & 16 \\
            $s$ & Stride of convolution & 8 \\
            $lr$ & Learning rate at the beginning & 0.001 \\
            \bottomrule
         \end{tabular}
        }
    \end{center}
\end{table*}

\textbf{GL-Transformer.} 
Due to the non-stationary nature of ship radiated noise, 
certain sampling points may experience abrupt changes. 
We attempted to optimize the full attention strategy and 
proposed a new attention mechanism that integrates local and global attention. 
The improved Transformer integrates Global Attention and Local Attention 
(referred to as GL-Transformer).

In the traditional Transformer model's attention layer, 
the similarity between $Query$ and $Key$ is determined solely 
based on their individual values, without fully incorporating 
contextual information. In our model, we address this limitation 
by leveraging the tensor obtained in the preceding step, 
which already captures the temporal locality and sparsity of the 
time series. We achieve this by initially convolving a sequence 
of contiguous sample points within the local vicinity to derive 
local trends for computing $Q$, $K$, and $V$. 
Subsequently, we establish associations between sparsely sampled 
points in the time series from a holistic standpoint to grasp the overall trend.

GL-Transformer first groups the spatial dimensions of features and calculates the attention within each group. 
Finally, it fuses the group attention globally to avoid query key matching that is irrelevant to local context. 
For local trends, we use a causal convolution with a kernel size of $k$ and a stride of 1 to transform the input into $Q$, $K$, and $V$. 
When $k=1$, it is the standard attention. For global trends, a convolution with a kernel size of 1 and a stride of s is used to convert the input into $Q$, $K$, and $V$. When $s=1$, this situation also degenerates into normative attention. 

Assuming $X$ is the input feature, $W^Q$, $W^K$, and $W^V$ are the weight matrices to be learned, the three parameters of one-dimensional convolution are the input feature, kernel size and stride, and $d_k$ represents the dimension of k. 
Taking local situations as an example, the calculation methods for $Q$, $K$, and $V$ are as follows:

\begin{equation}
    Q_{local} = Conv1D(X, k, 1) W^Q
\end{equation}

\begin{equation}
    K_{local} = Conv1D(X, k, 1) W^K 
\end{equation}

\begin{equation}
    V_{local} = Conv1D(X, k, 1) W^V
\end{equation}

The calculation of attention after fusion is as follows, where  $W_f$ represents the weight matrix and $b_f$ is the bias:

\begin{equation}
    Attention_{l} = softmax(\dfrac{Q_{local}  K_{local}^T} {\sqrt{d_k}}) V_{local}
\end{equation}

\begin{equation}
    Attention_{g} = softmax(\dfrac{Q_{global} K_{global}^T} {\sqrt{d_k}}) V_{global}
\end{equation}

\begin{align}
    Attention=sigmoid(W_f[Attention_{l};Attention_{g}]+b_f)
\end{align}

\section{Evaluation}

\subsection{Dataset and parameter settings}

To better validate the separation accuracy of the proposed network, we utilized authentic ship radiation noise data from the ShipsEar \cite{b27} and DeepShip datasets \cite{b28}. 
For the Deepship dataset, we selected radiation noise from four categories of ships: oil tankers, tugboats, passenger ships, and cargo ships. 
For the ShipsEar dataset, it includes four types of ship radiated noise and background noise recorded on the water surface.
Initially, all signals from each category were divided into segments of 2 seconds in length. 
For each dataset, the signals of different classes are additive mixed to obtain a total of 4096 mixed audio.
These mixed audio was split into training, validation, and test sets in a ratio of 7:2:1. The test set was utilized for model evaluation. The separation of mixed radiation noise data from pairs of the four ship categories was measured to assess separation effectiveness.

Regarding the network parameters, we set the number of epochs to 30 and the initial learning rate to 0.001. If the loss value did not decrease after 5 consecutive epochs, the learning rate was reduced to 0.0001. Additionally, for the dual-path network, we set the repeat parameter for the dual path to 6 and the number of heads for multi-head attention to 4. The key model parameters and their descriptions are recorded in Table 1.

\subsection{Evaluation metrics and Comparison models}
In our study on underwater acoustic signal separation, 
we sought to assess the efficacy of signal separation before 
and after the process. To achieve this, we employed 
three distinct metrics as evaluation criteria: 
signal-to-noise ratio (SNR), segmented signal-to-noise ratio (SegSNR), and scale invariant source to noise ratio improvement (SISNRi)  \cite{b29, b30}.  
A higher value indicates a more pronounced signal relative to noise, 
signifying a more effective separation outcome.

Segmented signal-to-noise ratio (SegSNR) is used to 
evaluate the frame level separation accuracy as well. 
In order to obtain segmented signal-to-noise ratio, 
it is necessary to divide the separated signal into several frames. 
For each frame of the signal, the signal-to-noise ratio is first calculated separately, and then the average signal-to-noise ratio of each frame is calculated. 

\begin{equation}
    \text{SegSNR} = \frac{1}{f_l}\sum_{i=0}^{f_l}\text{SNR}_{frame}(i)
\end{equation}

where $f_l$ denotes the number of frames and $\text{SNR}_{frame}$ denotes the SNR value of each frame:

\begin{equation}
    \text{SNR}_{frame}(i)=10log_{10}(\frac{\sum_{j=0}^{M_s-1}n^2(M_s-j)}{\sum_{j=0}^{M_s-1}[x(M_s-j)-x^*(M_s-j)]^2})
\end{equation}

where $M_s$ represents the number of samples per frame, $x$ represents the estimated signal, $x^*$ is a clean truth source signal.

The scale invariant signal-to-noise ratio improvement (SISNRi) 
is achieved by comparing the difference in scale invariant signal-to-noise ratio (SISNR) before and after signal separation,
where $X_E$ is an independent noise signal perpendicular to the estimated signal, 
$X_T$ is the true signal component in the estimated signal:

\begin{equation}
    \text{SISNR} = 10log_{10}\frac{||x_T||^2}{||x_E||^2}
\end{equation}

\begin{equation}
    x_T = \frac{x^*x}{||x||^2}x
\end{equation}

\begin{equation}
    x_E = x-x_T
\end{equation}

SISNRi is obtained by calculating the difference in SISNR before and after separation using the separation model:

\begin{equation}
    \text{SISNRi}=\text{SISNR}_{after}-\text{SISNR}_{before}
\end{equation}

Due to the fact that the separated $\text{SISNR}_{after}$ should be greater 
than the pre separated $\text{SISNR}_{before}$, SISNRi is usually a positive
value, indicating that the separation model has achieved a positive separation effect, 
and the larger the SISNRi value, the more effective the separation.

\begin{figure*}[htbp]
    \centerline{\includegraphics[width=1.2\linewidth]{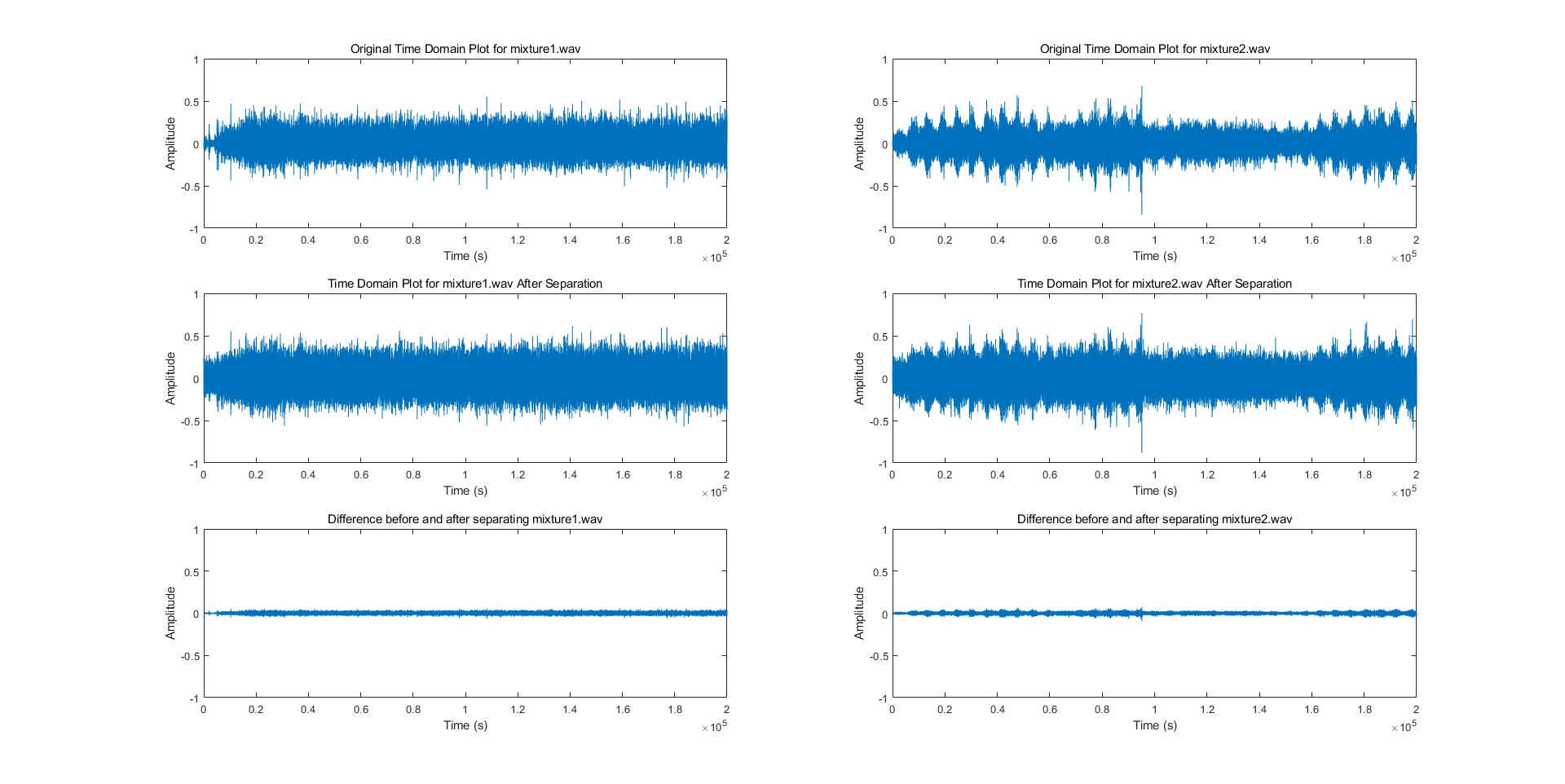}}
    \caption{The separation results are visualized in the form of time-domain waveforms. The two images on the top row represent the original audio. The second line shows the separation results obtained from the proposed model(Indiformer). The last line shows the absolute difference between the pure signal and its corresponding separated signal.}
    \label{fig}
\end{figure*}

We chose to compare UNet \cite{b18}, Res-UNet \cite{b20}, Conv-TasNet \cite{b24}, 
DPRNN \cite{b19}, and Mossformer \cite{b26}, 
which have shown strong performance in separating speech signals in recent years. 
Adaptive Filtering\cite{b4} and FastICA\cite{b15} were also juxtaposed for comparison, given their status as classical, traditional means of signal separation. 
To ensure consistency, we kept the epoch and 
initial learning rate the same for all models in our experiment. 
Additionally, for the comparison between DPRNN and Mossformer, 
we maintained an equal number of n\_repeat iterations.

\subsection{Validation and Performance Analysis}

\begin{table*}[htbp]
    \renewcommand{\arraystretch}{1.2}
    \caption{After mixing pairwise the four classes of data in \textbf{ShipsEar} and then separating them, the scores evaluated by Indiformer}
    \begin{center}
        \scalebox{1.2}{
        \begin{tabular}{|c|c|c|c|c|c|c|c|c|c|c|c|c|}
            \hline
            \multirow{2}{*}{Mixture} & \multicolumn{2}{c|}{(0,1)} & \multicolumn{2}{c|}{(0,2)} & \multicolumn{2}{c|}{(0,3)} & \multicolumn{2}{c|}{(1,2)} & \multicolumn{2}{c|}{(1,3)} & \multicolumn{2}{c|}{(2,3)} \\
            \cline{2-13}
            & 0 & 1 & 0 & 2 & 0 & 3 & 1 & 2 & 1 & 3 & 2 & 3 \\
            \hline
            SNR &19.31 & 18.22 & 18.67 & 19.10 & 17.89 & 18.43 & 18.76 & 18.98 & 19.55 & 18.01 & 17.68 & 19.42\\
            \hline
            SegSNR & 17.26 & 17.09 & 17.10 & 17.27 & 17.04 & 17.22 & 17.19 & 17.35 & 17.09 & 17.43 & 17.38 & 17.58 \\
            \hline
            SISNRi & 7.98 & 7.56 & 7.22 & 7.54 & 7.80 & 7.44 & 7.07 & 7.39 & 7.75 & 8.07 & 7.70 & 7.26 \\
            \hline
            \end{tabular}
        }
    \label{tab1}
    \end{center}
\end{table*}

\begin{table*}[htbp]
    \renewcommand{\arraystretch}{1.2}
    \caption{After mixing pairwise the four classes of data in \textbf{Deepship} and then separating them, the scores evaluated by Indiformer}
    \begin{center}
        \scalebox{1.2}{
        \begin{tabular}{|c|c|c|c|c|c|c|c|c|c|c|c|c|}
            \hline
            \multirow{2}{*}{Mixture} & \multicolumn{2}{c|}{(0,1)} & \multicolumn{2}{c|}{(0,2)} & \multicolumn{2}{c|}{(0,3)} & \multicolumn{2}{c|}{(1,2)} & \multicolumn{2}{c|}{(1,3)} & \multicolumn{2}{c|}{(2,3)} \\
            \cline{2-13}
            & 0 & 1 & 0 & 2 & 0 & 3 & 1 & 2 & 1 & 3 & 2 & 3 \\
            \hline
            SNR & 18.09 & 18.26 & 18.12 & 18.29 & 18.15 & 18.32 & 18.18 & 18.35 & 18.21 & 18.38 & 18.24 & 18.41\\
            \hline
            SegSNR & 17.92 & 17.78 & 17.96 & 17.82 & 18.00 & 17.86 & 18.04 & 17.90 & 18.08 & 17.94 & 18.12 & 17.98 \\
            \hline
            SISNRi & 7.87 & 7.82 & 7.77 & 7.72 & 7.67 & 7.62 & 7.57 & 7.52 & 7.47 & 7.42 & 7.37 & 7.32 \\
            \hline
            \end{tabular}
        }
    \label{tab1}
    \end{center}
\end{table*}

\begin{table*}[htbp]
    \renewcommand{\arraystretch}{1.2}
    \setlength{\tabcolsep}{9pt}
    \caption{Comparison of the proposed method with  Adaptive Filtering, Unet, ResUnet, Conv TasNet, DPRNN, and Mossformer in terms of model size, SNR, SegSNR, and SISNRi}
    \begin{center}
        \scalebox{1.2}{
        \begin{tabular}{cccccc}
            \toprule
            \textbf{Methods} & \textbf{Dataset} &\textbf{Model Size(M)} & \textbf{SNR(dB)} & \textbf{SegSNR(dB)} & \textbf{SISNRi(dB)} \\
            \midrule
            Adaptive Filtering \cite{b4}& \multirow{8}{*}{ShipsEar}&/ & 7.59 & 3.10 & 0.91 \\
            FastICA  \cite{b15} && / & 15.14 & 4.67 & 4.54 \\
            UNet \cite{b18} && 103.0 & 12.35 & 5.28 & 2.74 \\
            Res-UNet \cite{b20} && 33.4 & 13.19 & 8.09 & 2.79 \\
            Conv-TasNet \cite{b24} && \underline{5.1} & 16.84 & 14.65 & 4.63 \\
            DPRNN \cite{b25} && \textbf{3.6} & 18.42 & 16.29 & 5.87 \\
            Mossformer \cite{b26}& & 10.8 & \underline{18.83} & \underline{17.17} & \underline{7.31} \\
            \textbf{Proposed Method (Indiformer)} && 10.4 & \textbf{18.90} & \textbf{17.62} & \textbf{7.56} \\
            \\
            Adaptive Filtering \cite{b4}& \multirow{8}{*}{DeepShip}&/ & 8.61 & 2.85 & 0.98 \\
            FastICA  \cite{b15} && / & 11.87 & 5.84 & 3.78 \\
            UNet \cite{b18} && 103.0 & 13.72 & 5.95 & 3.02 \\
            Res-UNet \cite{b20} && 33.4 & 13.84 & 9.19 & 3.19 \\
            Conv-TasNet \cite{b24} && \underline{5.1} & 16.18 & 14.97 & 5.38 \\
            DPRNN \cite{b25} && \textbf{3.6} & 17.89 & 17.21 & 7.20 \\
            Mossformer \cite{b26}& & 10.8 & \underline{18.19} & \underline{17.73} & \underline{7.83} \\
            \textbf{Proposed Method (Indiformer)} && \underline{10.4} & \textbf{18.22} & \textbf{17.84} & \textbf{7.92} \\
            \bottomrule
         \end{tabular}
        }
    \end{center}
\end{table*}

In order to validate the effectiveness of the proposed model, a random segment of audio from the test set
was selected for mixing. This mixed audio was then input into the model, resulting in separated outputs. 
The test results are depicted in Figure 3. The audio is visualized in the form of time-domain waveforms. 
The top row of two images represents the original audio. 
The second row demonstrates the separated results obtained from the proposed model. 
The final row illustrates the absolute difference between the pure signals and their 
corresponding separated signals.
Upon comparison, it is observed that the two separated signals obtained from the network are essentially 
consistent with the time-domain waveforms of their respective target pure signals. 
To further validate this, the difference between the pure signals and their corresponding separated 
signals was calculated. The results show that the difference fluctuates around zero, 
thereby substantiating that the proposed model can successfully separate mixed speech and achieve 
commendable separation performance.

\begin{figure*}[htbp]
    \centerline{\includegraphics[width=18.5cm,height=8.3cm]{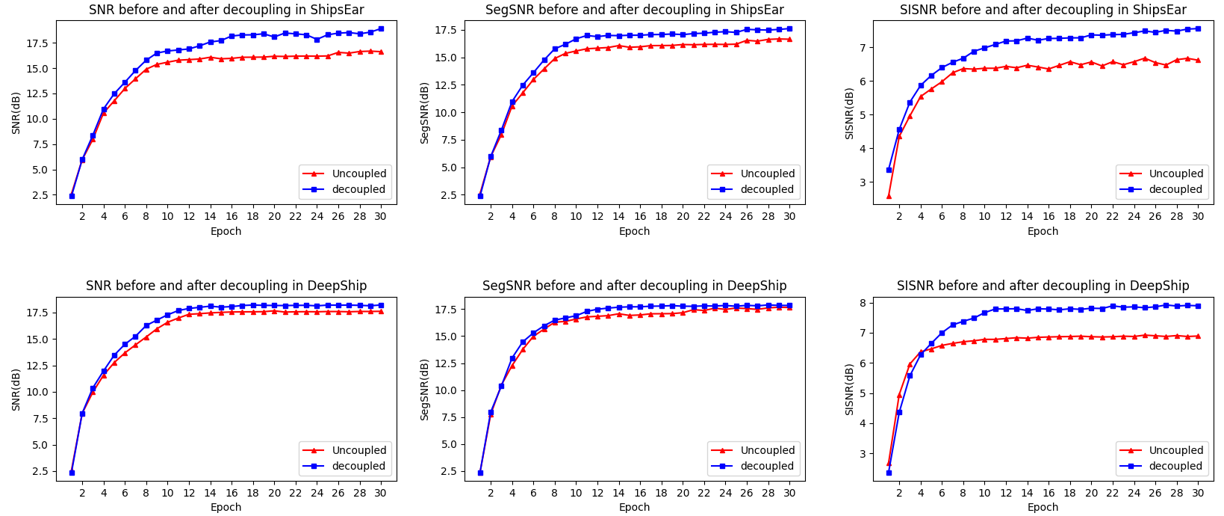}}
    \caption{ The ablation experiments conducted on the feature decoupling part were tested after 30 epochs of training.}
    \label{fig}
\end{figure*}

To evaluate the performance of our proposed method compared to other common separation models, 
we conducted a comparative analysis. 
Tables \uppercase\expandafter{\romannumeral2} and \uppercase\expandafter{\romannumeral3} show the mixed object separation experiments conducted on the ShipsEar dataset and
 Deepship dataset, respectively, using SNR, SegSNR, and SISNRi as scoring criteria. 
 In Table \uppercase\expandafter{\romannumeral2}, 0, 1, 2, and 3 respectively represent the four types of ships in the ShipsEar dataset:
  Fishboat, Sailboat, Piloship, and Roro. In Table \uppercase\expandafter{\romannumeral3}, 0, 1, 2, and 3 respectively represent the 
  Oil Tankers in the Deepship dataset, Tugboats, Passenger Ships,  Mix every two types of signals
   with the four types of cargo ships, for example (0,1) represents the mixed signal composed of 
   Fishboat and Sailboat, and so on for other labels. 
   The specific meaning in the table is to separate the target signal from the mixed signal of
    (a, b) as the evaluation score for Class A and Class B.

Table \uppercase\expandafter{\romannumeral4} summarizes the objective evaluations of these models. 
Using metrics such as SNR, SegSNR, and SISNRi, 
we tested the separation performance of different models on 
mixed signals, with all scores derived from averaging tests on 
mixed data segments from the ShipsEar test set. 
The results indicate that our method, Indiformer, 
has a smaller model size compared to Unet, ResUnet, and Mossformer. 
While slightly larger than Conv-TasNet and DPRNN, 
it is capable of handling tasks involving long signals. 
In terms of SNR, Indiformer performs closely to Mossformer and 
notably outperforms other methods. For SegSNR, Indiformer leads by a significant margin over several methods and still surpasses Conv-TasNet and DPRNN. 
Lastly, in terms of the SISNRi metric, Indiformer shows a notable improvement over other methods, confirming the superiority of our proposed approach based on these conclusions.

To verify the effectiveness of the feature decoupling module, 
we conducted ablation experiments on it. In the experiment, 
the proposed model and the featureless decoupling method that only
 includes the dual-path GL-Transformer were used to process the data in the same way, 
 dividing the training set, validation set, and test set. 
 The measurement indicators obtained were statistically analyzed and compared. 
 The performance of the two methods before and after decoupling on SNR, SegSNR, 
 and SISNRi after 30 epochs of training is shown in Figure 4, 
 where the red triangle represents the score without decoupling module, 
 and the blue rectangle represents the score of Indiformer on the three indicators. 
 After incorporating feature decoupling, there was an overall improvement 
 in performance across three key metrics. This suggests that initiating feature separation 
 on mixed signals will have a positive promoting effect on subsequent separation tasks.

\section{Conclusion}
In addressing the challenge of separating and reconstructing underwater passive ship radiated noise, we introduce an approach called Indiformer, building upon the foundation of the classical dual-path recurrent neural network. This method retains a dual-path architecture to effectively handle long and non-stationary underwater passive ship radiated noise. Our proposed technique involves decoupling features before separation, mapping reshaped tensor blocks into a space with more independent features. Additionally, we integrate the dual-path structure with local and global attention mechanisms, calculating local convolutions within chunks and equidistant global convolutions across chunks.

To evaluate the efficacy of our model, we conducted comparisons with several mainstream signal separation models using the ShipsEar dataset. Indiformer demonstrates robust separation capabilities, outperforming other methods to varying degrees across metrics like SNR, SegSNR, and SISNRi. 

In summary, our approach exhibits promising performance in the task of separating passive underwater acoustic signals, showing potential for applications in signal separation within underwater environmental engineering and military operations.

\balance
\begin{IEEEbiography}[{\includegraphics[width=1in,height=1.25in,clip,keepaspectratio]{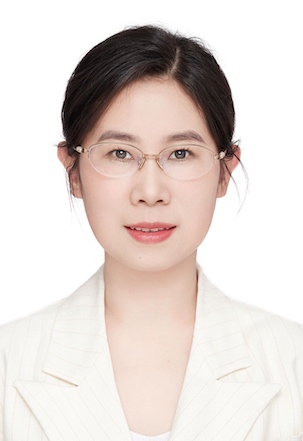}}]{}
    \textbf{Longyu Jiang} received her bachelor's, master's, and doctoral degrees from Wuhan University, Southeast University, and Grenoble Alpes University (France), respectively. She is currently a Distinguished Young Professor and doctoral supervisor at Southeast University. Her research primarily focuses on underwater acoustic signal and image processing, artificial intelligence, and big data. In recent years, she has led or participated in several key research projects, including China’s Key Special Projects, the National Natural Science Foundation of China, the French National Research Agency (ANR) Fund, and the China Scholarship Council’s Returned Scholar Fund. She has published over 30 papers in internationally renowned journals and conferences, such as IEEE Journal of Oceanic Engineering and The Journal of the Acoustical Society of America. Her major professional affiliations include serving as an evaluation expert for the China Scholarship Council’s government-sponsored study abroad programs, a committee member of the Visual Sensing Specialized Committee of the China Society of Image and Graphics, and a committee member of the Underwater Communication Specialized Committee of the Jiangsu Communication Association.
\end{IEEEbiography}

\begin{IEEEbiography}[{\includegraphics[width=1in,height=1.25in,clip,keepaspectratio]{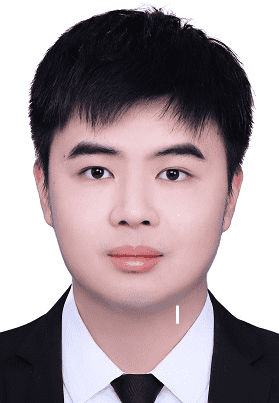}}]{}
    \textbf{Yucheng Liu} is currently pursuing a master's degree at the Joint Institute of Southeast University and Monash University. He earned his bachelor's degree from Shandong University. He is dedicated to research in underwater acoustic signal processing and has participated in multiple collaborative research projects between the Smart Ocean Laboratory of Southeast University and other institutions. His research interests include signal denoising, feature extraction, and deep learning.
\end{IEEEbiography}

\end{document}